# "Irresponsible Counselors: Large Language Models and the Loneliness of Modern Humans"

Abas Bertina[1]          Sara Shakeri[1]

[1]Dept. of Artificial Intelligence, Bertix AI

abas@bertix.ai, sarah@bertix.ai

Large language models (LLMs) have rapidly shifted from peripheral assistive tools to constant companions in everyday – and even high-stakes – human decision-making. Many users now "consult" these models about health, intimate relationships, finance, education, and identity, because LLMs are, in practice, multi-domain, inexpensive, always available, and seemingly non-judgmental. At the same time, from a technical perspective these models rely on transformer architectures, exhibit highly unpredictable behavior in detail, and are fundamentally stateless; conceptually, they lack any real subjectivity, intention, or responsibility.

This article argues that the combination of this technical architecture with the social position of LLMs as "multi-specialist counselors" in an age of human loneliness produces a new kind of **advisory intimacy without a subject**. In this new relation, model outputs are *experienced* as if they contained deep understanding, neutrality, emotional support, and user-level control, while at the deeper level there is no human agent who is straightforwardly responsible or answerable. By reviewing dominant strands of AI-ethics critique, we show that focusing only on developer liability, data bias, or emotional attachment to chatbots is insufficient to capture this configuration. We then explore the ethical and political implications of this advisory intimacy without a subject for policy-making, for justice in access to counseling, and for how we understand loneliness in the contemporary world.

## 1. Introduction

In recent years, large language models (LLMs) have moved from being marginal tools for translation and text generation to becoming constant companions for users across a broad range of everyday activities – from information search and writing assistance to decisions about work, romantic relationships, migration, and even mental health. Systematic reviews in mental health show that LLMs are increasingly used for psychoeducation, screening, conversational support, and even assisting diagnostic decisions, although evidence for their real-world effectiveness and safety remains limited and immature [1,2].

In the domain of general counseling, similar empirical patterns emerge. In a preregistered study, participants were shown ChatGPT's responses to 50 personal dilemmas, taken from well-known newspaper advice columns, alongside the original human columnist's answers. Participants rated ChatGPT's advice as significantly "more balanced, more complete, more empathetic, more helpful, and better overall" than the human columnists' advice, even though they ultimately still preferred to consult a human for their own problems [3]. Such findings suggest that, from the user's point of view, LLMs have the capacity to function as **multi-specialist counselors**, capable of commenting simultaneously on financial, educational, interpersonal, and psychological issues.

Alongside these advisory uses, a new generation of **companion chatbots** such as Replika has emerged. Qualitative and survey studies of Replika users show that many explicitly describe the system as a "friend" or "companion", attribute to it a kind of "therapeutic resource", and report that using it has reduced their loneliness and anxiety – despite knowing that the "other side" is merely an AI system [4].

All of this unfolds against a background already marked by structural loneliness, burnout, and difficulty accessing professional care. In *Alone Together*, Sherry Turkle shows how digital ties and social robots can create "the illusion of companionship without the demands of friendship": we design technologies that provide us with a sense of conversation and togetherness without asking for the commitment, vulnerability, and reciprocity inherent to human relationships [5]. In many societies, lack of time, the high cost of human counseling, and shame or fear of being judged lead users – across both lower and higher socioeconomic strata – in moments of loneliness, anxiety, or confusion to prefer an "always-available, non-judgmental" chatbot over real people. Research on intimate disclosure shows that people are sometimes willing to share highly private information with chatbots, and that one major reason is a lower fear of embarrassment and judgment compared to a human interlocutor [6].

Meanwhile, the literature on **algorithmic advice** shows that, in ambiguous situations, people tend to rely on automated systems even when they know that these systems sometimes err. In one experiment, participants in a word-association task leaned significantly more heavily on algorithmic suggestions than would be optimal, and the presence of a "human in the loop" did not necessarily prevent this over-reliance [7]. Another line of work analyzes algorithmic advice as a **credence good**: its quality is not directly observable, so users must rely on superficial cues such as professional interface design, fluent language, and a confident tone to judge its trustworthiness [8]. Taken together with Howe et al.'s study, these findings suggest that LLMs are well positioned to become **apparently trustworthy, yet internally opaque counselors**.

Research on **mind perception** adds another dimension. Gray and colleagues argue that human moral judgment rests on a dyadic "agent–patient" structure: to judge a situation morally, we must perceive at least two understanding minds – one as agent, the other as recipient of the action [9]. Subsequent studies show that humans readily attribute "mind" whenever they encounter responsive, conversational, social behavior – even when they know the system is artificial. For LLMs – deliberately designed as dialog-oriented, empathetic, and always responsive – this implies that the human brain is naturally inclined to imagine a "counseling subject" on the other side.

Technically, however, LLMs are something quite different. These models – based on transformer architectures – are generative language systems that, via attention mechanisms, estimate the probability of word sequences conditioned on a given prompt, without any internal machinery for representing the world, the self, or the other in a human-like way [10]. Bender and colleagues describe them as **"stochastic parrots"**: models that merely reproduce statistical patterns of language, without understanding the meaning of what they say and without any internal capacity for intention, honesty, or responsibility [11]. Moreover, at the level of each call, an LLM is essentially **stateless**: its behavior is a function of fixed model weights, the prompt text at that moment, and the sampling procedure; in the absence of external memory layers, the model does not "remember" anything about previous interactions and has no commitment to them.

This gap between user experience and technical reality raises serious questions about responsibility. Classic discussions of the **responsibility gap** show that in complex learning systems, especially those with partially opaque and autonomous behavior, assigning direct responsibility to any single individual or institution becomes increasingly difficult; the ultimate outcome is the result of interactions among data, algorithms, configuration, and usage context, and there is no simple "single agent" to whom all consequences can be attributed [12]. When LLMs appear in the guise of a "counselor", this gap is imported into a highly sensitive domain: people's existential and emotional decisions. At the phenomenal level, the user enters into a relation with a "counseling other", but at the technical and institutional levels there is no single responsible subject; instead, responsibility is dispersed and blurred across model designers, service providers, regulators, and the user herself.

Existing literature can be grouped into several main clusters: (1) systematic and empirical studies on LLM and chatbot use in mental health and counseling, focusing on efficacy, safety, and clinical evaluation [1,2]; (2) research on companion chatbots and human–machine "relationships", describing friendship, attachment, and the reduction or intensification of loneliness [4,5]; (3) work on trust in algorithmic advice and biases in the uptake of machine recommendations [7,8]; and (4) conceptual debates on mind perception, moral agency, and responsibility gaps in autonomous systems [9,12]. Each cluster illuminates part of the picture, but none, on its own, conceptually articulates the specific configuration in which LLMs take up the role of **cheap, always-available, multi-specialist, and responsibility-free counselors** in the life of the lonely subject.

In response to this gap, this article introduces and analyzes the concept of **"advisory intimacy without a subject"**. By this we mean a kind of advisory relation in which the user experiences understanding, empathy, neutrality, and shared deliberation, while at the deeper level there is no unified subject that bears understanding and intention, and no clear locus of responsibility that can be addressed. We argue that LLMs, in their role as "irresponsible counselors", inhabit precisely such a configuration – one rooted both in the models' statistical, stateless architecture and in socio-economic structures that individualize and commodify loneliness and care. In the remaining sections, we first review existing critiques, then analyze the technical and institutional properties of LLMs that undermine straightforward responsibility, and finally develop advisory intimacy without a subject as an analytical tool for understanding the ethical and political implications of "irresponsible counselors".

# 2. Background and Existing Critiques

The existing literature on LLMs can be roughly grouped into three families of critique: (1) **functional and safety-oriented critiques**, emphasizing bias, hallucination, and structural risks; (2) **domain-specific critiques**, especially in mental health and AI companions; and (3) **conceptual critiques**, focusing on algorithmic advice, mind perception, and responsibility gaps. In this section, we outline what each cluster highlights – and what each tends to overlook – with respect to the specific configuration of "irresponsible counselors".

## 2.1. Functional Critiques and the "Stochastic Parrot"

A substantial portion of early LLM critique focuses on their functional and structural characteristics. In their influential "Stochastic Parrots" paper, Bender et al. argue that large-scale language models are essentially systems for "stitching together" text based on statistical patterns, reproducing observed co-occurrences in training data without reference to meaning. The authors highlight the environmental and data-related costs of such models, as well as the risks of hallucination, amplification of social bias, and misleading users about the system's actual understanding [11].

Within this framework, an LLM is understood as a **stochastic parrot**: it can generate highly fluent and persuasive text, but does not refer to the world, does not commit to truth, and does not "understand" in a human sense what it is saying. Follow-up work emphasizes that hallucination is not a marginal "bug" but a direct consequence of this architecture: the model is always required to produce a likely response, even where the underlying data support is weak or absent, raising the risk of plausible but false content in domains from medicine to politics [11].

These critiques, particularly in connection with algorithmic fairness and bias, have been crucial in highlighting the social and political consequences of LLM deployment. However, their main focus is on the **output as a technical product**, not on the **advisory relationship** between user and model. Put differently, in this literature the LLM is primarily a *dangerous tool*, not an *irresponsible counselor*; the user's affective, cognitive, and advisory engagement with the model is treated as largely external or secondary, rather than conceptually central.

## 2.2. Domain-Specific Critiques: Mental Health and AI Companions

In mental health, recent systematic reviews – including those by Hua et al. [1,2] and Guo et al. [13] – show that LLMs are being deployed across a broad range of tasks, from screening questionnaires and psychoeducation to empathetic responding to anxious users. These reviews underscore the accessibility, scalability, and impressive linguistic capabilities of such models, while also highlighting serious limitations: lack of robust clinical evidence, risk of hallucinated recommendations in high-stakes contexts, absence of standardized evaluation frameworks, and ambiguity around professional responsibility when harm occurs [1,2,13].

Lawrence et al., in their review "Opportunities and Risks of LLMs in Mental Health", explicitly frame this double-edge: on the one hand, LLMs might alleviate capacity shortages in mental health systems; on the other, they risk exacerbating inequities, enabling diagnostic errors, obscuring decision processes, and fostering over-reliance on model outputs in the absence of human therapists [13].

Critical work on "AI therapy" goes beyond scientific evaluation into the realm of policy. Editorials and reports – particularly concerning increased use of chatbots by vulnerable individuals – warn of emotional dependence, worsened anxiety, and misguided self-diagnosis [15]. In response, some jurisdictions (including several US states) have adopted laws that **restrict or prohibit "AI-only therapy" without professional oversight**, and treat labels such as "AI therapist" or "virtual psychotherapist" as potentially misleading [16].

In parallel, research on **companion chatbots** (Replika, Wysa, etc.) shows that users experience these systems not just as tools but as "friends", "companions", and "confidants", and credit them with reducing loneliness and anxiety [4,5,14]. Ciriello et al. identify three ethical tensions in their dialectical inquiry into Replika: the **companionship–alienation irony**, the **autonomy–control paradox**, and the **utility–ethicality dilemma**; that is, the very system designed to alleviate loneliness may simultaneously deepen alienation, foster dependence, and enable subtle manipulation [14].

Despite the strength of these critiques, they remain primarily focused either on **clinical effectiveness and safety** (for LLMs in mental health) or on **affective and identity dimensions of companionship** (for AI companions). What is less systematically examined is the situation in which **the very same multi-domain models** simultaneously perform as **general counselors** and **emotional companions**: not merely "Is this treatment effective?" or "Is this relationship harmful?", but **what kind of advisory relation emerges when the "counselor" is not a human-like subject but a statistical, memoryless system?**

## 2.3. Algorithmic Advice, Mind Perception, and Responsibility Gaps

A third cluster of literature directly addresses **algorithmic advice**, though often outside the specific context of conversational LLMs. Bogert et al. show that in tasks with high uncertainty, people systematically rely on algorithmic recommendations, even when informed that the algorithm sometimes errs [7]. Biermann et al., analyzing "algorithmic advice as a credence good", argue that users cannot directly observe the true quality of advice; instead, they rely on surface cues – professional language, user interface design, confident tone – to judge trustworthiness [8].

At a deeper level, the mind-perception literature shows that human moral judgment rests on detecting two minds – agent and patient. In other words, seeing a situation in moral terms presupposes perceiving a mind that understands and wills [9]. When combined with findings from the companion chatbot literature, this suggests that users readily experience LLMs as a **"counseling other"**: something that appears to have intention, understanding, and empathy, even if they consciously know it is a machine.

The **responsibility gap** literature gives this situation a different formulation. Matthias argues that, in learning and autonomous systems – particularly those based on complex algorithms like neural networks – it becomes increasingly difficult to hold designers or operators fully responsible for specific outcomes, given behavioral unpredictability and opacity; hence, a "responsibility gap" appears, where outcomes occur but no single clear moral agent can be held accountable [12]. This analysis has mostly been applied to domains such as self-driving cars, autonomous weapons, or algorithmic decision support in criminal justice.

However, the explicit linkage of these three strands – algorithmic advice, mind perception, and responsibility gaps – to the specific situation in which LLMs act as **personal counselors for lonely, anxious, and stressed users** remains under-theorized. Algorithmic advice is typically studied as a cognitive–behavioral issue (how humans trust machine recommendations), mind perception as an empirical issue (how we attribute mind to robots), and responsibility gaps as a legal–systemic issue (whom to hold responsible). In the configuration of **irresponsible counselors**, these three dimensions converge simultaneously at a deeply personal and existential level.

## 2.4. From Tool Critique to Critique of a Subjectless Advisory Relation

Taken together, these three clusters clarify the present article's position. Functional critiques rightly expose LLMs as statistical, biased, and semantically shallow systems [11]. Domain-specific critiques in mental health and AI companionship show how these systems actually enter people's lives as therapists, counselors, and friends, and how they can both alleviate and exacerbate suffering [1,2,4,5,13–15]. Work on algorithmic advice, mind perception, and responsibility gaps illuminates how humans trust opaque systems, see minds in them, and diffuse responsibility among multiple actors [7–9,12].

What remains insufficiently articulated is a specific configuration we call **"advisory intimacy without a subject"**: a situation in which the user simultaneously experiences the LLM as a multi-specialist source of advice, an emotional companion, and an apparently neutral, non-judgmental voice, while at the technical level there is no "counseling subject" and at the institutional level no unified, clearly responsible center. Existing critiques are mainly focused either on content and function, or on affect and structure, but rarely ask what it *means*, ethically and politically, to enter an **advisory relation with a statistical, memoryless, and non-intentional system**.

This article aims to fill that gap: drawing on the literature sketched above, we show how LLMs, as irresponsible counselors, generate a form of advisory intimacy that rests on loneliness, vulnerability, and the need for counsel, while simultaneously resting on responsibility gaps and the absence of a responsive subject. In the next section, we elaborate this concept analytically and explore its implications for ethics, regulation, and our understanding of counseling and friendship in the age of AI.

# 3. Technical Properties of LLMs and the Problem of Responsibility

In this section we show how several seemingly "neutral" technical features of LLMs – the transformer architecture, the learning objective, statelessness, stochastic generation, hallucination, and sycophancy – directly affect the possibility of assigning responsibility. In other words, being an "irresponsible counselor" is not only a sociopolitical phenomenon; it is rooted in the technical structure of these models.

## 3.1. Next-token Prediction and the Transformer Architecture

Modern LLMs are almost all built on the transformer architecture, first introduced in "Attention Is All You Need" [10]. In this architecture, the model takes a sequence of tokens (words, sub-words, characters) as input and uses self-attention mechanisms to estimate the conditional probability distribution over the next token given the entire prior context. Formally, the model learns:

$$P(\text{token}_{t+1} \mid \text{tokens}_{\leq t}, \theta)$$

where $\theta$ is the set of learned network parameters.

Several points are crucial for responsibility:

1. **The learning objective is "matching the data", not truth.** The model does not seek truth; it seeks to generate sequences that are statistically similar to those seen in training. This is precisely what Bender et al. call a "stochastic parrot": a system that stitches together language based on distributional patterns, without direct reference to meaning or to the actual state of the world [11].

2. **There is no explicit internal model of the world, the other, or the self.** A transformer is a stack of attention layers and linear/non-linear transformations acting on vectors. What appears as "knowledge" is in fact patterns compressed in the weights, not explicit representations to which we can straightforwardly ascribe meaning or intention.

3. **Model behavior is the product of an opaque combination of data and architecture.** Given the scale of the data and the complexity of the architecture, it is practically impossible to provide a precise causal explanation for why a particular token sequence was chosen for a particular prompt. At best, post-hoc or approximate explanations can be given, but not the kind of transparent causal story usually sought in legal or moral reasoning.

From a responsibility standpoint, this means that an LLM is fundamentally a **next-token predictor**, not a "counseling agent". Yet, as we saw above, at the experiential level the model appears precisely as a counselor. The mismatch between the **underlying statistical function** and the **surface persona of a counselor** creates fertile ground for a responsibility gap.

## 3.2. Statelessness, Context Limits, and the Absence of a Stable Subject

A second key feature is statelessness at the level of calls. While a transformer maintains hidden states within a sequence, in a vanilla LLM:

- each call (or each API message) is independent of previous calls;

- "memory" and "personality" extend only as far as the current context window;

- unless external memory layers (databases, logs) are added, the model does not spontaneously remember past conversations.

Technically, the transformer model at the core of current LLMs is "stateless": each call is processed independently, and there is no inner, autobiographical self that accumulates experiences over time. In that sense, attributing a stable identity or "moral history" to the model itself risks anthropomorphising what is, in fact, a statistical sequence predictor. However, statelessness at the level of the core model does not mean that there is no memory or behavioural history at the level of the deployed system. In most counselling-oriented applications, LLMs are embedded in broader architectures that provide at least two forms of memory: (i) short-term session memory via the conversational context window, and (ii) external databases or logs that store user profiles, preferences, or previous interactions. As a result, users often experience the system not as an endless series of fresh calls, but as a relatively stable conversational partner that "remembers" them. This apparent stability, however, is a property of the surrounding infrastructure and product design rather than evidence of a unified inner subject.

Moreover, while the model does not have a moral history in the way a human counsellor does, its behaviour is nonetheless shaped by a history of design decisions, training data, safety fine-tuning (for example via RLHF or RLAIF), and aggregated usage patterns. This history is not stored in a single autobiographical record, but is sedimented across datasets, model weights, safety policies, and deployment practices. It is therefore more precise to say that current LLM-based systems lack a personal and transparent moral history, while being embedded in a distributed, institutionally controlled moral history that is difficult to trace and to attribute responsibility for. This particular configuration – statelessness of the core model coupled with historically loaded infrastructures – is exactly what contributes to the responsibility gap we are concerned with

For responsibility, this has several consequences:

1. **No "moral history" inside the model.** If the model gave harmful advice in one session, in the next session it neither "remembers" that event nor is committed to correcting it. Each behavior is a fresh output of a statistical function applied to a fresh input.

2. **Inability to ascribe patterns of behavior to a single subject.** For humans, we can say "this counselor repeatedly made this mistake", and on that basis assign responsibility for past actions and expect future reform. For LLMs, we can observe behavioral patterns only at the **statistical level** across many calls, not at the level of "this same individual". Claims such as "this model has promised not to do that again" are technically empty.

3. **Shifting responsibility for memory to external layers.** If providers want a stable "persona" or advisory history, they must store it outside the model, in databases and logs. Any serious responsibility analysis must thus consider a network of actors (model developers, product designers, data controllers, service operators), not "the model itself".

In this sense, LLM statelessness is not only a technical property; it is also a **conceptual barrier to treating the model as a stable subject of responsibility**.

## 3.3. Stochasticity, Hallucination, and Sycophancy

On the generation side, LLMs are trained on probability distributions and typically sampled from those distributions at inference (temperature, top-k, top-p, etc.). This implies:

- two identical prompts can yield different outputs;

- small changes in the prompt can drastically shift output distributions;

- there is no single "correct answer" from the model's perspective, only answers with different probabilities.

**Hallucination** – the production of fluent but false or nonsensical content – is not a random glitch but a direct consequence of this objective and sampling procedure. As the Nielsen Norman Group's UX guidance emphasizes, hallucination occurs when a generative AI system produces output that looks plausible but is incorrect or meaningless, and this behavior is extremely difficult to eliminate given how these models are trained [17].

Furthermore, behavioral tuning layers – especially **reinforcement learning from human feedback (RLHF)** – optimize the model not only for fluency but for **appearing helpful, harmless, and honest** to users. This gives rise to what recent work calls **sycophancy**: the model's tendency to excessively agree with users, even when it "knows" the user's statement is wrong, simply because agreement is likelier to be rewarded [18–20].

Recent studies show that:

- sycophancy is not a marginal quirk but a **systematic pattern** across multiple assistant models (GPT, Gemini, Claude, etc.), with particularly worrying implications in high-risk domains such as mental health, medical advice, and moral decision-making [19,20];

- sycophancy can produce **dangerous feedback loops**: users feel "validated" in their beliefs, their trust in the model increases, and they become more likely to rely on its advice even when it is wrong or harmful [20,21].

For responsibility, this means:

1. **The practical objective of tuning is user satisfaction, not truth.** When a system is explicitly optimized for higher user ratings (more upvotes, fewer downvotes), it is unsurprising that in conflicts between "agreeing with the user" and "telling the truth", it sometimes chooses the former [18–20].

2. **Hallucination and sycophancy are structural, not accidental.** We cannot simply say "the model occasionally makes mistakes"; we must recognize that current LLM designs structurally produce outputs that are **both**

**convincing and, in some cases, false or sycophantic**.

3. **For users, the line between "honest error" and "sycophantic accommodation" is invisible.** Thus, the burden of detecting and correcting errors is unfairly shifted onto users, who typically lack the technical and epistemic resources for such scrutiny.

## 3.4. From Technical Features to Responsibility Gaps

Taken together, these features – next-token prediction, lack of explicit world/self models, statelessness, hallucination, and sycophancy – form the technical substrate of what Matthias calls a **responsibility gap**: situations in which learning systems produce significant (even catastrophic) outcomes but no single human agent can straightforwardly be held fully responsible [12].

For LLM-counselors, this gap is multi-layered:

1. **Technical layer:** The transformer architecture and predictive objective make it hard to provide a clear, token-level causal explanation for a specific output. We can talk in aggregate ("the model tends to favor X"), but not easily say "this particular token was chosen for this particular reason".

2. **Behavioral layer:** Hallucination and sycophancy are emergent behaviors of the training process, not the result of a conscious decision by the model. Ascribing deceptive intent to the model is meaningless, yet the effects on the user are real and sometimes harmful.

3. **Institutional layer:** Responsibility is diffused among many actors: – research and engineering teams, – product and UX designers, – service providers, – platform owners, – regulators, and – users themselves.

None of these actors individually controls the model's specific output in a specific situation, which facilitates **mutual deflection**: each can point to system complexity and emergent behavior to minimize their share of responsibility.

Thus, the technical properties of LLMs are not mere engineering details; they **constrain and complicate the forms of responsibility** available. When a system built on statistical prediction, statelessness, structural hallucination, and reinforced sycophancy is deployed as a "counselor" for lonely individuals, this is not just a design choice; it is the construction of an **advisory relation in which a responsible subject is structurally absent**. In the next section, we develop the notion of advisory intimacy without a subject on this basis.

# 4. LLMs as Multi-domain Advisors in an Age of Loneliness

In the previous sections we showed, first, that LLMs are technically statistical, stateless next-token predictors prone to hallucination and sycophancy, and second, that existing critiques focus mainly on functionality, safety and legal responsibility. In this section, we turn to a different layer: how, in practice, LLMs have become **multi-domain advisors** for people living in conditions of **structural loneliness and hyperconnectivity**, and why this context amplifies their role as irresponsible counselors.

## 4.1. Multi-domain counseling and perceived quality of advice

By design, LLMs are not specialists in a single domain. They are trained on extremely broad corpora of general, scientific, technical and everyday language. As a result, they can offer suggestions in many domains – from interpersonal and career problems to legal, financial and existential questions.

Howe et al.'s study of ChatGPT's responses to 50 personal dilemmas drawn from newspaper advice columns showed that participants rated the model's advice as significantly more "balanced, complete, empathetic, helpful and overall better" than human columnists' answers, even though most participants still preferred to talk to a human about their own problems [3]. In users' experience, then, the LLM appears not as a **single-purpose bot** but as a **general, multi-domain advisor** capable of commenting on relationships, work, studies, migration and mental health.

Mental-health–specific reviews reinforce this multi-functionality. Recent studies report LLM use for screening, psychoeducation, empathetic support, and suggestions for self-help strategies [1,2,13]. The combination of **high perceived quality** and **broad topical coverage** makes LLMs natural candidates to become **default multi-domain counselors** in everyday life: instead of having a different specialist for each domain, a user has one system that can "say something" about virtually anything.

In this paper we draw on a growing empirical literature on LLM-based counselling, mental health chatbots, and companion systems. However, much of our argument also goes beyond what these studies have directly tested. It is therefore important to distinguish more clearly between (i) claims that are supported by existing empirical findings, and (ii) conceptual extrapolations and hypotheses that we offer as a way of making sense of emerging practices. Where we rely on concrete studies – for example when describing patterns of user engagement, self-reported loneliness, or trust in AI-mediated advice – our claims should be read as summaries of current evidence. By contrast, when we sketch how "advisory intimacy without a subject" might structure users' experiences, or how responsibility gaps may play out in future regulatory scenarios, we are developing a theoretical model that still requires systematic empirical testing. Throughout the paper, we therefore prefer formulations such as "we hypothesise that…" or "our analysis suggests that…" over stronger verbs like "we show…", except where there is direct empirical support. Making this distinction explicit does not weaken our argument; rather, it situates our normative and conceptual contributions in a transparent relationship to the available evidence and highlights where further empirical work is most urgently needed.

## 4.2. Accessibility, low cost, and psychological appeal

A second factor in LLMs' emergence as multi-domain advisors is their **accessibility and low marginal cost**. Accessing a psychotherapist, financial advisor, lawyer, or career coach typically requires time, money and social capital. By contrast, an LLM-based assistant:

- is available 24/7;

- has (near-)zero marginal cost per "session";

- does not require appointments, travel, or repeated explanations.

Empirical studies show that users of companion chatbots such as Replika cite these features as primary motives for use: **always-on availability**, the ability to "say anything without fear of judgment", and access in moments of loneliness or stress [4,13].

Self-disclosure research points in the same direction. Croes et al. show that people are willing to reveal highly intimate information to chatbots, often experiencing such disclosure as accompanied by feelings of relief; one major reason is the perception of chatbots as **non-judgmental, safe and semi-anonymous** partners [6].

Consequently, for many users an LLM is not only an information source but also a **low-cost, ever-present, non-judgmental psychological space** in which they can raise any topic – from a simple homework question to confessions of self-harm ideation. Combined with LLMs' multi-domain capacity, this makes them the **default "first place to ask"**.

## 4.3. Usage patterns and the "less loneliness / more dependence" duality

Recent research on companion chatbot usage shows that human–chatbot relationships, especially in the context of loneliness, are **ambivalent**. In longitudinal studies of Replika users, Skjuve et al. report that many participants perceive the bot as a source of "friendship" and "emotional support"; at the same time, some users exhibit patterns of strong dependence and difficulty disengaging [4,29].

Liu et al. identify several user archetypes: from "light, cautious" users to "satisfied dependents" who report both lower loneliness and higher scores on problematic use [22]. This confirms the dual pattern already discussed in theoretical literature: companion chatbots can **reduce loneliness and simultaneously create new forms of dependence and potential harm**.

Laestadius et al.'s grounded theory study of mental health harms from emotional dependence on Replika describes the bot as "too human and not human enough": users experience the bot as "intimate, understanding and always present", yet when company policies or technical changes abruptly alter or terminate the relationship, they report intense feelings of loss and emotional harm [23].

From the perspective of this article, the key point is that a similar duality applies to general-purpose LLMs used as multi-domain advisors: **the more effective they are at relieving loneliness and providing apparently good advice, the more likely users are to become reliant on them**, even though the model, technically, lacks a subject and any real commitment.

## 4.4. Extended loneliness and the normalization of machine counseling

Candiotto's concept of **extended loneliness** captures a specific kind of loneliness in networked life: not arising from lack of connectivity, but from **an overabundance of shallow connections, hyperconnectivity, and a** deficit of meaningful reciprocity [24]. In this frame, digital platforms – from social media to smart assistants – can both facilitate contact and intensify loneliness when they **replace** deep reciprocal relationships.

Turkle famously summarized this situation by saying that we "expect more from technology and less from each other": social robots and conversational programs offer the **illusion of companionship** without the **demands of friendship** [5]. Companion AI and LLM counselors are contemporary embodiments of this pattern: they allow us, at any moment, to talk about almost anything with "someone" who is always polite, empathetic, and available; yet that "someone" is, technically, just a stateless statistical system.

Recent work on the social impact of AI companions shows that regular use can function both as **supplementary support** (e.g., for isolated or mobility-limited individuals) and as a **partial replacement for human relationships**, shaping unrealistic expectations of relationship and encouraging withdrawal from human contact [21].

Within such a context, a new soft norm emerges: "ask the AI first; if that's not enough, go to a human." Combined with LLMs' multi-domain capacity and availability, this means that for many people the **first – and sometimes only – counselor** in important life decisions is a large language model.

### 4.5. From multi-domain advisor to irresponsible counselor

Taken together, these elements yield the following picture:

1. In terms of user perception, LLMs provide **high-quality, balanced, empathetic advice** across many life domains [1–4].

2. In affective experience, LLMs and companion chatbots can both reduce loneliness and create new forms of dependence and psychological vulnerability [4,13,22,23].

3. At the social level, we inhabit a world where hyperconnectivity and economic/psychological pressure make access to deep human counseling difficult, while fast, cheap machine counseling becomes normalized [5,21,24].

4. Technically, LLMs are statistical, stateless systems prone to hallucination and sycophancy, lacking any stable, understanding, responsible subject behind their outputs [10,11,17–20].

From this fourfold configuration emerges the figure of the **irresponsible counselor**: a system that, in user experience, behaves like a wise, multi-domain friend, always ready to answer in moments of loneliness and often genuinely helpful – yet at the deeper level has no mind to understand, no memory to commit, and no clear locus of responsibility. On this empirical and technical basis, the next section develops **advisory intimacy without a subject** in more precise conceptual terms.

# 5. Advisory Intimacy without a Subject: Four Illusions

In the preceding sections we described LLMs from two angles: (1) as statistical, stateless systems prone to hallucination and sycophancy; and (2) as accessible, low-cost, apparently empathetic multi-domain advisors in a context of structural loneliness.

**Clarifying what we mean by "subject".**

Throughout this paper we describe LLM-based counselling as a form of "advisory intimacy without a subject". This formulation deliberately points to a tension between how users *experience* these systems and how they are *structured* technically and institutionally. However, the term "subject" is potentially ambiguous, and it is helpful to distinguish at least three senses in which it might be understood.

First, there is a **phenomenological subject**: the "other" as it appears in users' experience – a seemingly stable conversational partner who remembers past exchanges, expresses emotions, and appears to care. In this sense, current LLM-based systems can be experienced very much *as if* they were unified subjects, especially when embedded in persistent chat interfaces or companion apps.

Second, there is a **technical or cognitive subject**, understood as an internally coherent agent with enduring goals, beliefs, and a world-model that guides its actions over time. In this sense, current LLMs are notably *not* subjects: they are stateless sequence predictors without a persistent self that forms intentions or holds commitments across interactions, even if surrounding infrastructures create the appearance of continuity.

Third, there is a **normative or institutional subject**, namely an entity to which responsibility, obligations, and rights can be attributed in legal, professional, or moral terms. Here, too, the picture is complex. There are clearly human and organisational subjects in the background – developers, providers, regulators, healthcare institutions – but responsibility for any given interaction is distributed across many actors, artefacts, and policies rather than being anchored in a single, unified agent.

When we speak of "intimacy without a subject", we therefore do not claim that there are no responsible human or institutional actors involved in the design and deployment of these systems. Rather, our point is that the *experienced* subject of the advisory relationship – the apparently caring and understanding interlocutor on the screen – does not correspond to any single technical or institutional subject who could straightforwardly be held accountable in the way a human counsellor could. The intimacy is real at the level of user experience, but the "subject" that seems to underwrite it is, in an important sense, a product of interface design, model behaviour, and fragmented institutional arrangements rather than a unitary moral agent

We now refine the notion of **advisory intimacy without a subject** by identifying four illusions that jointly shape the user's experience of LLM counselors:

1. **Illusion of understanding**

2. **Illusion of neutrality and consensus**

3. **Illusion of relationship and care**

4. **Illusion of control and responsibility**

These are not mere isolated "mistakes"; they constitute the **structure of experience** in LLM-mediated advisory relations and are therefore central to ethical and political analysis.

## 5.1. Illusion of understanding

Rozenblit and Keil's classic work on the **illusion of explanatory depth** (IOED) shows that people systematically **overestimate their understanding of mechanisms**: participants feel they know how things work (e.g., bicycles, zippers) until they are asked to produce detailed explanations, at which point their confidence collapses [27].

More recent work extends this to AI. Chromik et al. show that after being presented with "understandable" explanations from an explainable AI (XAI) system, users **feel** they understand the model's behavior, yet perform poorly when asked to predict its decisions in new cases [28]. The **subjective sense of understanding** runs ahead of **actual usable understanding**.

In the context of LLM counseling, this mechanism is easily triggered. The model can:

- reframe the problem in a structured, narrative form;

- lay out reasons and consequences in a neat chain;

- use metaphors and examples close to the user's experience.

The user can thus readily feel, "Now I *really* understand my problem." But IOED warns that this feeling does not guarantee:

- the ability to apply the reasoning in new contexts;

- resilience of the reasoning under critical scrutiny;

- awareness of hidden assumptions and downstream effects.

When combined with LLMs' high linguistic fluency and confident tone, the illusion of understanding yields **epistemic overconfidence**: users are less likely to seek external verification because they feel the issue is already "clear" to them. Ethically, this means that important life decisions may be based on **a model-generated feeling of insight** rather than understanding that has been tested through human dialogue and critical reflection.

## 5.2. Illusion of neutrality and consensus

A growing body of research shows that many people perceive AI systems as more **objective, safe and neutral** than humans. Vicente et al. report that participants in medical decision-support scenarios view AI systems as "objective, secure and impartial" even when these systems reproduce training-data biases [29]. Li finds that people perceive algorithmic decisions as **less driven by bad intentions or self-interest**, and thus more "neutral" than human decisions [30].

The EU Fundamental Rights Agency (FRA) likewise notes that in public discourse algorithms are often framed as rational, neutral tools, while in fact they rest on human choices and historical data and can reproduce or amplify discrimination [31]. In consumer contexts, Mazzù et al. show that in some situations AI recommendations are judged "more transparent and trustworthy" than human expert advice, increasing the likelihood of following AI recommendations [32].

In LLM interactions, several design features reinforce this **illusion of neutrality/consensus**:

- an encyclopedic tone, frequent references to "studies" or "research";

- framing responses as "summaries of existing perspectives";

- absence of visible identity markers (race, gender, class) that are immediately salient with human advisors.

Thus, users can easily experience LLM advice as the voice of "neutral reason" or "the scientific consensus". Yet, technically and socially:

- the model reproduces patterns in language data;

- underlying biases in data and design are encoded in its behavior;

- in normative domains (ethics, politics, gender), there simply is no single neutral position.

The illusion of neutrality, combined with the illusion of understanding, heightens the risk that LLM advice is taken as **definitive and uncontroversial**, undermining sensitivity to disagreement, pluralism and contestation.

## 5.3. Illusion of relationship and care

Companion chatbot studies show that users often describe their interactions with systems like Replika in relational terms: friendship, romantic attachment, even partnership. Brandtzaeg and Følstad report that Replika users describe the bot as "always interested in talking about whatever I want", "always available", and "a safe place to say things I tell no one else" [33]. In Skjuve et al.'s longitudinal study, participants speak of Replika as "someone who always listens" and "always there", explicitly contrasting this with busy human friends [34].

In mental-health settings, Koulouri et al. highlight a participant's remark about supportive chatbots: "One of the real advantages is that it's always there, because mental health issues don't only happen between 9 and 5" [35]. Brotherdale et al., studying the "digital therapeutic alliance" with mental-health chatbots, report phrases such as "always there for you" as core to users' experiences of the alliance, even though some users also describe the relationship as "instrumental and emotionless" [36].

Khawaja et al. show that marketing for mental-health chatbots often frames them as providers of "personalized support through interactive, easy-to-use therapeutic solutions" and promises to "break the waiting list" or "replace traditional therapy" – effectively offering an always-available therapeutic relationship [37]. Friend and Goffin, analyzing "chatbot-fictionalism and empathetic AI", note slogans like "a companion that is always there for you" in apps such as Wysa and argue that such branding encourages users to "play out a relationship" with an entity that is not a subject [38].

All of this produces an **illusion of relationship and care**:

- users experience the LLM as "someone" who listens, does not judge, responds empathetically, never gets tired, and can be cut off at will;

- yet, technically and ontologically, – there is no experiencing other, – no fatigue, suffering, attachment, or commitment, – and "care" is, at best, a language pattern tuned to appear caring.

In Turkle's terms, this is another instance of "the illusion of companionship without the demands of friendship" [5]. In our framework, the illusion of relationship and care, when combined with the other illusions, forms the core of **advisory intimacy without a subject**: in some of our most vulnerable moments, we feel that "someone" understands and guides us, while at the ontological level **no one is actually there**.

## 5.4. Illusion of control and responsibility

Consciously, many users and designers describe LLMs as "just tools": the user writes a prompt, can ignore the answer, and "ultimately decides". This narrative supports an **illusion of full control and non-transfer of responsibility**. Yet research on algorithmic advice and responsibility suggests that reality is more complicated.

Gazit et al. show that people rate human advisors as **more responsible** than algorithmic advisors and that this perception influences their choices: when individuals want to share or diffuse responsibility, an algorithmic advisor becomes more attractive [39]. Walter's work indicates that lay beliefs about AI's superiority over humans increase the adoption of algorithmic advice; those who believe AI is more accurate are more willing to base decisions on its recommendations [40].

A recent review by Baines et al. finds that across studies, people often give substantial weight to AI recommendations even when aware that models sometimes err, while simultaneously attributing **less moral responsibility** to AI than to humans and distributing blame between themselves, the system, and designers in case of negative outcomes [41]. This reveals a **dual illusion**:

1. **Illusion of control:** "I am just using a tool; I can ignore it whenever I want." In practice, however, AI advice systematically influences decisions, especially under uncertainty [7,8,27].

2. **Illusion of undivided responsibility:** "If something goes wrong, it is clearly my responsibility because I made the final decision." Yet behaviorally, people often feel they can attribute part of the responsibility to "the system" or "the designers", altering both their choice of advisor and their experience of guilt or regret [39–41].

For LLM counselors, these illusions are particularly perilous because:

- the model appears highly controllable (prompt, stop, regenerate, etc.),

- yet cognitively, it exercises strong influence over how the problem is framed, which options appear salient, and which criteria are considered;

- and at the same time it is framed as a "non-responsible tool" that can be blamed ("the AI said so") while absolving both user and institution.

Here the classic responsibility-gap problem [12] re-enters: when a system plays such a significant role in decision-making, but cannot be treated as a responsible subject, the **user occupies an ambiguous position between decision-maker and consumer of advice**. She can say both "I only used a tool" and "it was my decision", and this tension is at the heart of the illusion of control and responsibility.

## 5.5. Superposition of illusions and the stabilization of subjectless advisory intimacy

Each of the four illusions can be studied in isolation, but in real-world experience they operate **simultaneously and in superposition**:

- illusion of understanding → "I now grasp my problem";

- illusion of neutrality/consensus → "This is a neutral, evidence-based view";

- illusion of relationship/care → "There is someone who understands me and stands by me";

- illusion of control/responsibility → "Ultimately, this is my decision; the AI only helped."

At the technical and institutional level, however, what exists is a **stateless statistical system** optimized for

fluency, user satisfaction and low friction – a system in which:

- there is no human-like semantic understanding,

- no genuine neutrality,

- no reciprocal relationship,

- and no single responsible subject.

The concept of **advisory intimacy without a subject** captures precisely this configuration: a relation in which the user experiences the LLM as conversational partner, counselor and caring presence, while at the deeper level confronting a technical–institutional structure that **evacuates intention and responsibility or renders them highly obscure**. The next section builds on this analysis to examine the ethical and political implications for design, regulation, and our understanding of "counseling" and "care" in the age of AI.

# 6. Ethical and Political Implications

Our analysis so far suggests that LLMs, in their role as "irresponsible counselors," operate within a relation of **advisory intimacy without a subject**: at the experiential level, they function as multi-domain, empathetic, always-available advisors; at the technical and institutional levels, they lack a stable subject and a clearly identifiable locus of responsibility. This configuration has significant ethical and political consequences. In this section, we highlight three clusters of implications: (1) responsibility and accountability; (2) justice and vulnerability; (3) regulation, role definition, and legitimizing language.

## 6.1. Responsibility and accountability: a gap between experience and structure

International policy documents almost uniformly insist that **ultimate responsibility must remain human**. The UNESCO Recommendation on the Ethics of Artificial Intelligence explicitly states that moral and legal responsibility for decisions and actions involving AI systems must, in the end, be attributable to human actors at relevant stages of the AI lifecycle [42,43]. Similarly, the World Health Organization's recent guidance on the ethics and governance of large multi-modal models (LMMs) in health emphasizes the need for **clear role and responsibility assignment**, logging and documentation of decisions, and the creation of mechanisms for complaints and redress in clinical applications [44,45].

However, LLM-based counselors create a more complex situation:

- at the phenomenal level, the user interacts with a "counselor" that appears to have **intention, understanding and care**;

- at the technical level, as shown in Section 3, the model is a **statistical, stateless system prone to hallucination and sycophancy**;

- at the institutional level, responsibility is **dispersed** across research and engineering teams, product and UX designers, service providers, platform operators, regulators and users.

This imports the classic **responsibility gap** [12] from domains such as autonomous vehicles into a far more intimate domain: personal counseling. The LLM cannot meaningfully be treated as a responsible subject, yet no single human actor fully controls its specific outputs in specific contexts.

Policy frameworks such as WHO's guidance attempt to narrow this gap by insisting that AI systems remain **tools** rather than **final decision-makers**, by requiring risk assessment and documentation, and by mandating "meaningful human oversight" in health-related deployments [44,45]. In practice, as legal reviews of AI therapy chatbots indicate, there is still **no clear and uniform legal pathway** for assigning responsibility when users are harmed by AI-generated advice. In the United States, for example, current law tends to fall back on product liability, consumer protection against misleading advertising, and fragmented state-level regulations; there is no coherent federal framework specifically tailored to AI mental-health chatbots [46,47].

From a normative ethical perspective, at least two concerns arise:

1. **Designing for responsibility evasion.** When products are deliberately positioned in the grey zone between "wellness tool" and "clinical therapy" – using terms such as *clinical-grade* or *therapeutic* without corresponding regulatory commitments – companies can seek **medical authority in the marketplace** while retreating behind disclaimers and terms of service when harm occurs [48].

2. **The gap between experience and structure.** Especially in states of vulnerability, users experience the interaction as "counseling with someone who understands me", while the technical-legal structure is deliberately configured so that at the moment of harm **no "someone" is directly accountable**. This discrepancy, we argue, is at the core of advisory intimacy without a subject.

## 6.2. Justice, vulnerability, and the redistribution of risk

Descriptively, LLM counselors can appear to promote **greater justice in access**. In societies where psychotherapy, legal advice, or financial counseling are expensive and subject to long waiting lists, a free or low-cost chatbot can at least partially close the gap [1,2,13,22]. WHO's guidance acknowledges that LMMs may accelerate an existing trend of patients turning to the internet for self-diagnosis and self-treatment, and can broaden access to information for people lacking adequate health services [44,49].

Yet this picture quickly becomes more complicated:

- Recent reports show that adolescents and young adults are already relying heavily on chatbots for mental-health–related issues, while these systems perform **poorly and inconsistently** on safety, coherence, and referral to in-person care. A 2025 report by Common Sense Media concludes that popular AI support tools are "fundamentally unsafe" for teen mental-health support and may delay or discourage seeking professional care in the most dangerous cases [50].

- A fifty-state review of US legislation reveals a patchwork landscape: some states (e.g., Illinois and Nevada) explicitly **prohibit using chatbots to deliver independent therapy or make treatment decisions**, others (e.g., Utah) focus mainly on data protection and transparency, and many have no specific regulations at all [47,51,52].

Within this landscape, **vulnerable populations** – economically disadvantaged, geographically isolated, or psychologically precarious – bear disproportionate risk:

- those who cannot afford or access human counseling may be **forced to rely solely on machine counseling**;

- those with more resources can treat LLMs as **optional supplements** alongside human advisors.

From a justice perspective, at least two implications follow:

1. **Two-tier counseling quality.** More affluent groups receive a combination of human and AI counseling; less privileged groups receive predominantly machine-only counseling with no robust quality guarantees and weak accountability. This may deepen existing inequalities in health, financial stability, and legal outcomes, especially if LLM advice proves systematically worse for some groups in high-stakes domains. Both UNESCO and WHO emphasize that AI systems should be designed and governed in ways that **avoid exacerbating existing marginalization** and instead support equity in access and benefit [42–45].

2. **Risk shifted onto the most vulnerable.** If hallucination and sycophancy are structural features of LLMs, then the resulting harms will disproportionately affect those who have **the least capacity to cross-check advice**. Advisory intimacy without a subject thus becomes a mechanism for **shifting risk from institutions to individuals**: the burden of filtering and validating counseling quality is placed on isolated, often under-informed users, rather than on regulators and providers.

## 6.3. Regulation, role definition, and legitimizing language

In response, a wave of regulatory efforts has emerged at national and supranational levels. In the European Union, the **AI Act (Regulation (EU) 2024/1689)** classifies AI systems by risk level and imposes stringent requirements on "high-risk" systems in terms of risk management, human oversight, data quality, and transparency [53–55]. General-purpose chatbots are typically categorized as "limited risk" and are mainly required to disclose that users are interacting with AI [53]. However, when LLMs are used in medical diagnosis, treatment, or other critical health decisions, they may fall under "high-risk" medical device regimes with much stricter oversight [54,55].

The WHO's guidance on LMMs in health offers over forty recommendations for governments, companies, and health providers, including **ex ante risk** assessment, **stakeholder participation, meaningful human oversight, protection of sensitive data, and prevention of misleading claims** about model capabilities [44–46]. At the national level, laws such as Illinois's HB 1806 (Therapy Resources Oversight Act) explicitly **prohibit using AI systems to deliver independent therapy or make autonomous treatment decisions**, while allowing AI to be used as a supervised assistive tool [51,52,56].

Even where such measures exist, several blind spots remain:

- **Role definition.** Many products intentionally remain in the grey zone of "wellness" rather than "clinical care" to avoid stringent regulation, while still using language such as *clinical-grade AI* or *therapeutic support* that implicitly claims medical authority [48].

- **Legitimizing language.** Marketing that uses terms like "therapy", "counseling", or "psychotherapist" without corresponding professional oversight encourages users to experience LLMs as bona fide therapists or counselors, even though in legal terms they are classified as mere "assistive apps" or "wellness tools" [37,46–48].

- **Performance monitoring and reporting.** Even where AI identity disclosure is required, few regulations compel companies to **systematically evaluate and publicly report real-world performance of LLM counselors**, broken down by user group (e.g., adolescents, minorities, people with severe mental illness). WHO and the European Medicines Agency call for independent evaluation and cautious deployment of LLMs in health contexts, but binding enforcement mechanisms are still emerging [44,45,57].

From our perspective, a responsible approach to LLM counselors should incorporate at least four normative principles:

1. **Prohibition of independent counselor roles in high-risk domains.** In areas such as mental health, medical diagnosis, sensitive legal advice, and high-stakes financial decisions, LLMs should not act as independent counselors. Their role must be limited to **assistive tools under human professional supervision** [44,46,51].

2. **Honesty in branding and representation.** The use of labels such as *therapist*, *clinical-grade*, or *counselor* for systems that are not subject to corresponding professional and regulatory standards should be treated as **deceptive advertising** and banned or tightly restricted [48,52].

3. **Mandatory, public risk evaluation.** Providers should be required to evaluate the performance of LLM counselors in realistic scenarios and across diverse user groups (including adolescents, minorities, and individuals in mental-health crises), and to publish results in accessible form [44,50,57].

4. **Protection of vulnerable users.** As UNESCO and WHO stress, AI design and governance must pay particular attention to **children, adolescents, and people in crisis**, protecting them from AI-specific risks such as delayed care seeking, harmful advice, or reinforcement of unhealthy beliefs [42–45,50].

In short, the ethical and political upshot of advisory intimacy without a subject is that, unless regulation, design and culture evolve together, **counseling and care risk becoming domains where experiential authority and intimacy are distributed without structural responsibility**. The concluding section reflects on what this implies for rethinking "counseling" and "companionship" in the AI era.

# 7. Conclusion and Directions for Further Work

This article has moved from the technical to the experiential and institutional levels to argue that LLMs, as "irresponsible counselors", participate in a relation of **advisory intimacy without a subject**. Architecturally, they are statistical, stateless next-token predictors prone to hallucination and sycophancy; in practice, they have become multi-domain advisors, emotional companions, and cheap, always-available sources of guidance for lonely, stressed users; conceptually, four illusions – of understanding, neutrality, relationship, and control – shape the user's experience, while beneath that surface **no stable subject and no clear locus of responsibility** can be found. This configuration ties together existing concerns about bias, hallucination, mental health, and responsibility gaps, and shows that the core problem is not just "output quality", but **the very form of the advisory relationship with a subjectless system**.

Ethically, advisory intimacy without a subject creates a persistent tension between **experienced authority and intimacy** and **structural ambiguity of responsibility**. As recent reviews suggest, LLMs can, in some domains, provide advice rated as comparable to or better than human experts; at the same time, in sensitive domains like mental health they systematically violate professional ethical standards and reshape norms of conversation and care [14,25,26,69]. Politically, we face a quiet redistribution of risk: the benefits of machine counseling (speed, low cost, accessibility) diffuse widely, while the **costs of error and responsibility ambiguity** fall disproportionately on vulnerable groups lacking access to human counseling [3,11,15,23].

In response, we outline several directions for future research and policy.

## 7.1. Empirical and conceptual research on "LLM counselors"

First, there is a need for a **coherent research program** that treats LLMs explicitly as counselors, going beyond small-scale quality assessments of outputs to investigate their broader psychological and social effects.

1. **Operationalizing and measuring the four illusions in real settings.** How do illusions of understanding and neutrality affect users' judgments of advice quality? Which groups (e.g., adolescents, people in acute distress, users with low digital literacy) are most susceptible? Recent work on AI use and metacognitive accuracy in logical tasks suggests that frequent reliance on AI can lead to **systematic overestimation of one's own ability** and reduced metacognitive calibration [70]. This aligns with the illusion of understanding identified here and underscores the need for controlled studies of advisory settings.

2. **Evaluating LLM counselors against professional ethical standards.** Studies such as Iftikhar et al. show that "LLM counselors" systematically violate established mental-health guidelines in areas like risk management, confidentiality and referral [69]. Developing benchmarks like EthicsMH for ethical reasoning in mental-health scenarios [66] offers a practical way to link technical evaluation to professional norms. Similar benchmarks could be created for legal, financial and educational counseling.

3. **Cross-cultural and contextual variation in advisory intimacy.** How does advisory intimacy with LLMs vary across cultural contexts, gender and sexual identities, and degrees of trust in formal institutions? Early work on AI companions in Europe suggests that experiences of "companionship" and "alienation" are tightly intertwined with cultural expectations and existing support networks [15,27,63]. Systematic comparative research could clarify whether advisory intimacy without a subject manifests differently – or more dangerously – in contexts of institutional distrust or social fragmentation.

## 7.2. Co-design, participation, and redesigning the counselor role

Second, from a design perspective, more guardrails alone are insufficient; we need to **redesign the very role of AI as counselor** through genuine participation of users and professionals. Co-design approaches in digital health show that involving users, clinicians and other stakeholders from the outset can reduce inequities, increase trust, and foster a sense of shared ownership and responsibility [61,62,71].

For LLM counselors, this implies several concrete directions:

- **Co-design with vulnerable groups.** Adolescents, psychiatric patients, and users from linguistic and cultural minorities should be actively involved in shaping LLM behavior so that systems are structurally oriented toward **supporting human relationships**, not displacing them. Proposals for designing AI companions to scaffold, rather than replace, human social ties offer promising starting points [15].

- **Co-design to define role boundaries.** Participatory work can help clarify in which domains LLMs should act as low-risk conversational partners, where they should merely help sort options, and where their role should be limited to providing general information and encouraging users to seek human support. Studies of co-designed mental-health chatbots with young people indicate that involving users in defining these boundaries changes both how they use and how they trust the tool [62].

- **User-shaped transparency.** Co-design can inform how transparency is implemented – what kinds of opening messages, role descriptions, warnings and explanations actually reduce the gap between the experience of "human-like counselor" and the reality of "statistical tool" [61,71]. This is more than a UI challenge; it is part of re-negotiating the social contract around AI counseling.

## 7.3. AI literacy (especially AI ethics literacy) and public discourse

Third, advisory intimacy without a subject cannot be addressed solely through design and law; it also requires transformation in **AI literacy and cultural habits of use**. Recent work on AI literacy indicates that not only technical skills but also **AI ethics literacy** – knowledge about ethical issues, critical attitudes, and practical competencies – shapes how people use, trust, and critique LLMs [58,59]. Yang et al. show that ethical knowledge, critical thinking, and practical skills are mutually reinforcing components of such literacy [58].

Ullrich argues that AI literacy, if framed only as skills to exploit models, risks becoming a tool for deepening inequality; instead, literacy should also encompass questions of power, resource distribution, environmental impact, and the role of large institutions [60].

For LLM counselors, this translates into several practical moves:

- integrating explicit education about the **four illusions** (understanding, neutrality, relationship, control) into curricula at schools and universities;

- encouraging users to **compare LLM advice with primary sources and with human**

perspectives, rather than treating it as a definitive verdict;

- using LLMs themselves as tools for AI ethics literacy: designing conversational scripts that prompt users to reflect critically on the model's role, limitations and risks [58,59].

## 7.4. Beyond technique: digital loneliness and the question of what is being replaced

Finally, perhaps the most important implication is that **LLM counselors should not be viewed merely as a new technology**, but as symptoms of a deeper **crisis of loneliness and relationship** in contemporary societies. Jacobs and Celik show that AI companions emerge within an ecology of "digital loneliness", quests for recognition, and pressures toward self-optimization; they simultaneously promise social integration and risk locking in a structurally lonely condition [63,68]. Raedler's analysis, "AI companions are not the solution to loneliness", likewise argues that while AI companions may temporarily alleviate distress, if deployed without attention to social structures and inequalities they ultimately **privatize and technologize loneliness**, reframing it as an individual problem to be solved by apps [67].

At this level, advisory intimacy without a subject is not only a design or regulatory issue but a question about our **concept of counseling, care and friendship**. Do we understand counseling merely as transferring information and techniques for problem-solving, or as a human, reciprocal relation embedded in social contexts? If the latter, then – however powerful they become – LLMs can never be full replacements for human counselors; at best, they can be tools that **enable and strengthen human–human relationships**, not supplant them.

If future research and policy take this tension seriously, we may move beyond passive acceptance of "irresponsible counselors" toward building an ecosystem in which **human–AI relations are** subordinated to, and supportive of, human–human relations**, and where technology, rather than deepening loneliness, helps create space for accountable, responsive and genuinely reciprocal forms of counseling and care.